\documentclass[preprint2]{aastex61}

\newcommand\aastex{AAS\TeX}

\received{January 20, 2019}
\revised{March 12, 2019}
\accepted{\today}
\submitjournal{ApJ}

\shorttitle{\aastex\ sample article}
\shortauthors{Rodighiero et al.}

\begin{document}

\title{AGN in dusty starbursts at \lowercase{$z$}=2: feedback still to kick in}

\correspondingauthor{G. Rodighiero}
\email{giulia.rodighiero@unipd.it}

\author{G. Rodighiero}
\affil{Dipartimento di Fisica e Astronomia, Universit\'a di Padova,
Vicolo dell'Osservatorio, 3 
35122 Padova, Italy}
\affil{INAF Osservatorio Astronomico di Padova, vicolo dell’Osservatorio 5, I-35122 Padova, Italy}
\author{A. Enia}
\affil{Dipartimento di Fisica e Astronomia, Universit\'a di Padova,
Vicolo dell'Osservatorio, 3 
35122 Padova, Italy}
\author{I. Delvecchio}
\affil{Laboratoire AIM-Paris-Saclay, CEA/DSM-CNRS-Université Paris Diderot, Irfu/Service d’Astrophysique, CEA-Saclay, Service d’Astrophysique,
F-91191 Gif-sur-Yvette, France}
\author{A. Lapi}
\affil{SISSA, Via Bonomea 265, 34136 Trieste, Italy}
\affil{IFPU - Institute for fundamental physics of the Universe, Via Beirut 2, 34014 Trieste, Italy}
\author{G. E. Magdis}
\affiliation{Cosmic Dawn Center at the Niels Bohr Institute, University of Copenhagen and DTU-Space, Technical University of Denmark}
\affiliation{Niels Bohr Institute, University of Copenhagen, Lyngbyvej 2, DK-2100 Copenhagen, Denmark}
\affiliation{Institute for Astronomy, Astrophysics, Space Applications and Remote Sensing, National Observatory of Athens, GR-15236 Athens,
Greece}

\author{W. Rujopakarn}
\affiliation{Department of Physics, Faculty of Science, Chulalongkorn University, 254 Phayathai Road, Pathumwan,
Bangkok 10330, Thailand}
\author{C. Mancini}
\affil{Dipartimento di Fisica e Astronomia, Universit\'a di Padova,
Vicolo dell'Osservatorio, 3 
35122 Padova, Italy}
\affil{INAF Osservatorio Astronomico di Padova, vicolo dell’Osservatorio 5, I-35122 Padova, Italy}
\author{L. Rodr\'{\i}guez-Mu\~noz}
\affil{Dipartimento di Fisica e Astronomia, Universit\'a di Padova,
Vicolo dell'Osservatorio, 3 
35122 Padova, Italy}
\author{R. Carraro}
\affil{Instituto de F\'\i{}sica y Astronom\'\i{}a, 
   			Universidad de Valpara\'\i{}so,
              Gran Bretana 1111, Playa Ancha, Valpara\'\i{}so, Chile}
\author{E. Iani}
\affil{Dipartimento di Fisica e Astronomia, Universit\'a di Padova,
Vicolo dell'Osservatorio, 3 
35122 Padova, Italy}
\author{M. Negrello}
\affil{School of Physics and Astronomy, Cardiff University, The Parade, Cardiff CF24 3AA, UK}
\author{A. Franceschini}
\affil{Dipartimento di Fisica e Astronomia, Universit\'a di Padova,
Vicolo dell'Osservatorio, 3 
35122 Padova, Italy}
\author{A. Renzini}
\affil{INAF Osservatorio Astronomico di Padova, vicolo dell’Osservatorio 5, I-35122 Padova, Italy}
\author{C. Gruppioni}
\affil{Osservatorio di Astrofisica e Scienza dello Spazio – Istituto Nazionale di Astrofisica, via Gobetti 93/3, I-40129, Bologna, Italy}
\author{M. Perna}
\affil{INAF - Osservatorio Astrofisico di Arcetri, Largo Enrico Fermi 5, I-50125 Firenze, Italy}
\author{I. Baronchelli}
\affil{Dipartimento di Fisica e Astronomia, Universit\'a di Padova,
Vicolo dell'Osservatorio, 3 
35122 Padova, Italy}
\author{A. Puglisi}
\affil{Laboratoire AIM-Paris-Saclay, CEA/DSM-CNRS-Université Paris Diderot, Irfu/Service d’Astrophysique, CEA-Saclay, Service d’Astrophysique,
F-91191 Gif-sur-Yvette, France}
\author{P. Cassata}
\affil{Dipartimento di Fisica e Astronomia, Universit\'a di Padova,
Vicolo dell'Osservatorio, 3 
35122 Padova, Italy}
\author{E. Daddi}
\affil{Laboratoire AIM-Paris-Saclay, CEA/DSM-CNRS-Université Paris Diderot, Irfu/Service d’Astrophysique, CEA-Saclay, Service d’Astrophysique,
F-91191 Gif-sur-Yvette, France}
\author{L. Morselli}
\affil{Dipartimento di Fisica e Astronomia, Universit\'a di Padova,
Vicolo dell'Osservatorio, 3 
35122 Padova, Italy}
\author{J. Silverman}
\affil{Kavli Institute for the Physics and Mathematics of the Universe (WPI), Todai Institutes for Advanced Study, the University of Tokyo, Kashiwanoha, Kashiwa,
277-8583, Japan}
\affil{Department of Astronomy, School of Science, The University of Tokyo, 7-3-1 Hongo, Bunkyo, Tokyo 113-0033, Japan}






\begin{abstract}
We investigate a sample of 152 dusty sources at $1.5<z<2.5$ to understand the connection of enhanced Star-Formation-Rate (SFR) and Black-Hole-Accretion-Rate (BHAR). The sources are {\it Herschel}-selected, having stellar masses 
$M_*>10^{10} M_{\odot}$ and SFR  ($\sim100-1000 M_\odot/yr$) elevated ($>4\times$) above the star-forming "main sequence", classifying them as Starbursts (SB).
Through a multiwavelength fitting approach (including a dusty torus component), we divided the sample into active SBs (dominated by an AGN emission, SBs-AGN, $\sim23\%$ of the sample) and purely star-forming SBs (SBs-SFR). We visually inspected their HST/UV-restframe maps: SBs-SFR are generally irregular and composite systems; $\sim50\%$ of SBs-AGN are instead dominated by regular compact morphologies. 
We then found archival ALMA continuum counterparts for 33 galaxies (12 SBs-AGN and 21 SBs-SFR). For these sources we computed dust masses, and, with standard assumptions, we also guessed total molecular gas-masses.
SBs turn to be gas rich systems ($f_{gas}=M_{gas}/(M_{gas}+M_*)\simeq 20\%-70\%$), and the gas fractions of the two SB classes are very similar ($f_{gas}=43\pm4\%$ and $f_{gas}=42\pm2\%$).

Our results show that  SBs are consistent with a mixture of:  
1) highly star-forming merging systems (dominating the SBs-SFR), and 
2) primordial galaxies, rapidly growing their $M_*$ together with their Black Hole (mainly the more compact SBs-AGN). 
Anyway, feedback effects have not reduced their $f_{gas}$ yet. 
Indeed, SBs at $z=2$, with relatively low bolometric AGN luminosities in the range $10^{44}<L_{bol}(AGN)<10^{46}$ erg/s (compared to bright optical and X-ray
quasars), are still relatively far from the epoch when the AGN feedback will quench the SFR in the host and will substantially depress the gas fractions. 

\end{abstract}




\section{Introduction}\label{sec:intro}
The existence of a natural correlation between stellar mass and SFR for the bulk of star forming galaxies \cite[i.e. the so called Main Sequence, hereafter MS,][]{2007ApJ...670..156D, 2010MNRAS.401.1521M, 2012ApJ...754L..29W, 2014ApJS..214...15S}
and the identification of highly star-forming outliers above it,
implies that two main processes of galaxy growth occur: 1) a secular growth along a (quasi) steady state, and 2) stochastic episodes of major galaxy growth, possibly driven by major mergers that trigger short-lived and intense starbursts (SB). Similarly, two different ways of Black Hole (BH) growth seem to hold. Likely, secular processes dominate the growth of intermediate-to-low luminosity BHs, e.g. through continuous gas refuelling
\citep{2012ApJ...753L..30M}. 
Instead, the most luminous population of AGN may experience a different growth history strictly connected to the starburst activity in off-MS galaxies. Indeed, SBs show the following indications of enhanced AGN activity: 
1) higher average X-ray luminosities \citep[][and Carraro et al., in prep]{2015ApJ...800L..10R} 
2) higher AGN fraction (up to 50-80$\%$) both in the local Universe \citep{2012NewAR..56...93A}, and at high-$z$ \citep{2012MNRAS.427.3103B, 2018arXiv180703378P}, compared to MS galaxies \cite[$\sim 25-30\%$,][]{2009A&A...507.1277B}
This implies that the AGN duty cycle is higher above the MS, and the black hole accretion rate (BHAR) more efficient.

Many observational efforts have been devoted to understand what triggers the fast gas consumption rate in these spectacular sources \citep{2015ApJ...812L..23S, 2018arXiv180900715C, 2019A&A...623A..64C}. Mergers are often invoked as the more likely mechanism, although it is still not clear if this is sufficient to explain the enhanced star formation of SBs compared to normal galaxies
\citep{2018ApJ...867...92S}.
Some authors suggest that
the merger mechanism enhancing SFR in many SBs could be responsible also for the AGN ignition \cite[e.g.][]{2005Natur.433..604D, 2015MNRAS.447.2123C, 2018arXiv180506902S}. This scenario is, however, still quite debated \cite[e.g.][]{2016ApJ...833..152M, 2019MNRAS.482.4454W}.



In this Letter, we investigate the properties of starbursts in COSMOS at $1.5<z<2.5$. We study their morphologies, AGN contribution and molecular gas contents (for a sub-sample with an ALMA continuum observation) to infer potential observational evidences of feedback induced by AGN.


We adopt a \cite{2003PASP..115..763C} IMF assuming a standard cosmology with $H_0=70$, $\Omega_\lambda=0.7$, $\Omega_0=0.3$.

\begin{figure*}
\centering
\includegraphics[width=15cm]{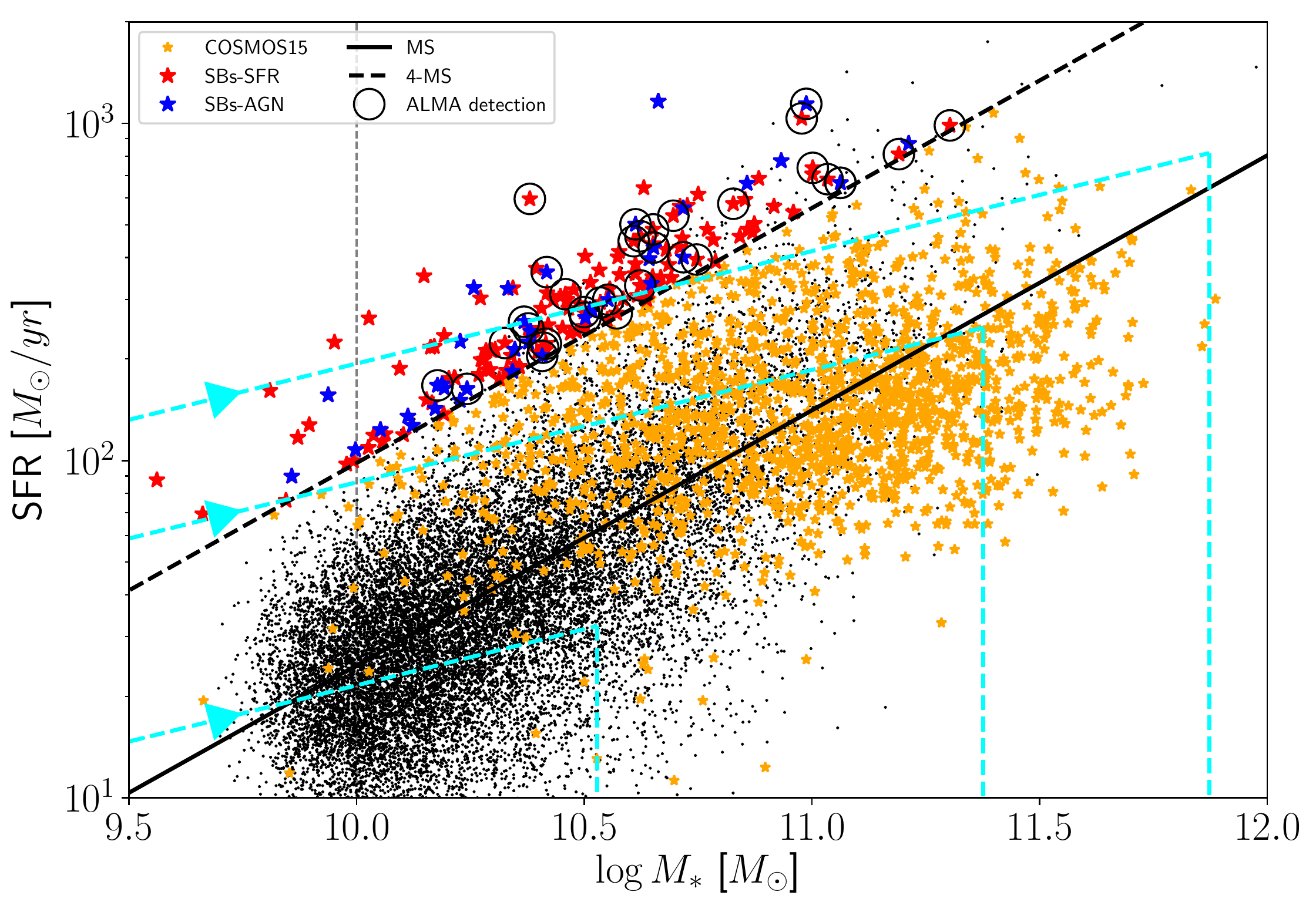}
\caption{
    Sample selection in the SFR-$M_*$ plane for the COSMOS field at $1.5<z<2.5$. Orange stars mark the original {\it Herschel} sample.
    Blue and red stars 
    show the position of SBs-AGN and SBs-SFR, respectively (see Section \ref{sec:sed}).
    Black open circles indicate the SB sources with an ALMA detection. Black dots are star-forming BzK 
\citep{2007ApJ...670..156D}. We report as dashed cyan lines evolutionary tracks of galaxies \citep{2018ApJ...857...22L} starting with different initial SFR and final total $M_*$ (time follows the arrows direction, see discussion in Section 6.2).
    The vertical black dashed line marks the mass limit of our SBs selection (see text for details).}
\label{fig:selection}
\end{figure*}

\section{Dusty starbursts at $1.5<z<2.5$: sample selection}
We adopt the {\it Herschel} far-Infrared catalog, associated to the COSMOS2015 sample \citep{2016ApJS..224...24L},
to select a sample of highly star-forming dusty galaxies at the peak of the cosmic SFR density (i.e. $1.5<z<2.5$).
PACS and SPIRE-{\it Herschel} data are originally from the PEP \citep{2011A&A...532A..90L} 
and HerMES 
\citep{2012MNRAS.424.1614O} 
surveys, with corresponding photometry extracted with a 24$\mu$m priors PSF fitting procedure.
 Sources identification, multiwavelength photometry, $M_*$ and redshifts are from the COSMOS2015 sample. Starbursts (SBs) are selected to have a SFR well elevated above the MS at $z=2$, at least a factor 4, as in \cite{2011ApJ...739L..40R}.


\subsection{SED fitting: SFR and AGN computation}\label{sec:sed}
We fit the multiwavelength SEDs of the $Herschel$ sources to derive their physical characterization.
We used both the \emph{MAGPHYS} \citep{2008MNRAS.388.1595D}, and the \emph{SED3FIT} \citep{2013A&A...551A.100B} SED-fitting codes, the latter accounting for an additional AGN component (dusty torus).
\emph{MAGPHYS} 
relies on the energy balance between the dust-absorbed stellar continuum and the reprocessed dust emission at infrared wavelengths. 
\emph{SED3FIT} combines the emission from stars, dust heated by star formation, and a possible AGN-torus component from the library of \cite{2012MNRAS.426..120F}. 
We fitted each observed SED by using the best available redshift (either spectroscopic or photometric) as input.
The SFR was derived from the total IR (rest frame [8-1000] $\mu$m) luminosity taken from the best-fit galaxy SED (subtracted by the AGN luminosity, if present), assuming a \cite{1998ARA&A..36..189K} conversion.

Out of 1790 {\it Herschel} detected sources at $1.5<z<2.5$, we identified a sample of 164 SBs (see Figure \ref{fig:selection}). We limit our analysis to $M_*>10^{10}M_\odot$ \citep[to ensure an unbiased mass complete selection, see][]{2016ApJS..224...24L} reducing the SB sample to 152 objects.

\subsection{AGN classification}
In order to quantify the relative incidence of a possible AGN component, we fitted each individual observed SED.
The fit obtained with the AGN is preferred if the reduced $\chi^2$ value of the best fit (at $\ge$99\% confidence level, on the basis of a Fisher test) is significantly smaller than that obtained from the fit without the AGN \cite[see][]{2014MNRAS.439.2736D}. From our analysis, 35 out of 152 starbursts (about 23\%) show a significant AGN component.
In the following, we will refer to these two classes as SBs with evidences of a nuclear activity (SBs-AGN) and SFR dominated SBs (SBs-SFR), respectively.
We have verified that out of 152 SBs in our starting sample, only eight are classified as X-ray AGN \citep{2016ApJS..224...24L}, six of which are identified by our AGN classification. This check ensures that we recover the bulk (i.e. 75\%) of the classical X-ray/AGN selection\footnote{As a consistency check, for X-ray undetected SBs-AGN, we computed the expected [2-10]keV X-ray fluxes, derived from the AGN bolometric luminosities obtained by the SED fitting procedure described in Section \ref{sec:sed}. By applying a suited bolometric correction \citep{2018MNRAS.475.1887Y}, and a K-correction=$(1+z)^{\Gamma-2}$ (with $\Gamma=1.5$, consistently with the stacked Hardness Ratios of our sources), we obtain a median F[2-10]keV on the order of $\sim1.7\times10^{-15}$ keV, right below the limit of the Chandra detectability in this band \cite[$\sim1.9\times10^{-15}$ keV,][]{2016ApJ...830..100M}}, extending the sample to include also the most obscured active sources \cite[e.g.][]{2012MNRAS.427.3103B, 2016MNRAS.458.4297G}. 

Moreover, since at $1.5<z<2.5$ the IRAC bands span the rest-frame near-IR where the galaxy stellar emission peaks, to ensure that the AGN classification is not contaminated we performed a further test. Thus, to verify if substantially old stellar populations could bias the AGN selection, we computed the mass weighted ages of all SBs in our sample with and without the dusty torus component: in spite of a significant scatter ($\sim$0.3dex), no offset is observed in the measured average stellar ages of the systems, both for SBs-SFR and SBs-AGN (probing that the {\em SED3FIT} software correctly recovers, in a statistical sense, the near-IR light arising from the stellar populations and from the torus).
We have also tested our AGN selection procedure against classical IRAC/\emph{Spitzer} colour-colour diagnostics, as the one proposed by \cite{2012ApJ...748..142D}, finding that $\sim80\%$ of the SBs-AGN fall into the AGN region, while just one of the SBs-SFR are mis-classified as AGN (see Fig.\,\ref{fig:agn_cutout}, left panel).
A comparison with the average fraction of the luminosities coming from the AGN component at [5-40]$\mu$m in the SED modelling, confirms that sources lying above the dashed line in the left panel of Fig.\,\ref{fig:agn_cutout} have four times larger AGN contributions (25\% vs 6\%, in terms of the light fraction arising from the torus in the mid-IR) with respect to the sources below the same line. 
This indicates that the procedure adopted in this work to classify AGN preferentially selects sources dominated by a torus component at mid-IR wavelengths (as expected). 
The comparison of our AGN selection criterium with the colour-colour plot presented in Fig. 2 (left panel) shows that our SBs-AGN sample is highly reliable\footnote{Even if starburst templates are a dominant contaminant of mid-IR AGN selection techniques \citep{2007A&A...470...21R, 2012ApJ...748..142D} have demonstrated that their conservative selection minimizes the inclusion of purely star-forming LIRGS and ULIRGS, whose templates 
begin to enter the IRAC selection region only at $z>2.7$.} (contamination $\le20\%$), but not necessarily complete \citep{2012ApJ...748..142D}.

\section{UV-restframe HST morphological analysis}
We performed a visual inspection on the COSMOS/HST ACS $i$-band images of the 140 (out of 152)  SBs for which the data are available (corresponding to the UV-restframe at $z=2$). 
Excluding undetected sources (9 out of 33 SBs-AGN, 39 out of 107 SBs-SFR), this analysis reveals that the two starburst classes have statistically different typical morphologies: SBs-SFR are disturbed systems (56 out of 68, i.e. $\sim$82$\pm11$\% of the HST detected sample), with evident tidal interaction between multiple components (possibly ongoing mergers or clumpy disks); SBs-AGN are (13 out of 24 detection, $\sim$55$\pm15$\%) dominated by regular compact and symmetric morphologies.
We note that for Type 1 sources the HST imaging would be dominated by the AGN outshining the hosts. From our SED fitting, we estimate that just a fraction of the compact SBs-AGN are consistent with a Type 1 classification, lowering to $\sim$34\%  the percentage of compact AGN in the sample, still significantly higher then the corresponding value among the SBs-SFR ($\sim$18\%). To overcome the uncertainties related to our SED fitting approach on the classification of Type 1 AGN, and the shallowness of the X-ray Chandra data in COSMOS, we rely on a stacking procedure to compute separately the average soft ($S$) and hard ($H$) X-ray fluxes for the different AGN morphological classes. 
We used the CSTACK tool\footnote{http://cstack.ucsd.edu/cstack/} (v3.1, T. Miyaji) and derive  the average Hardness Ratio, $HR=(H-S)/(H+S)$, of our SBs-AGN sample
\cite[see details on the procedure in][]{2015ApJ...800L..10R}.
We found that compact, $i$-band undetected and extended sources have compatible $HR=-0.12\pm0.11$, $0.02\pm0.31$, $-0.04\pm0.13$, respectively. This result suggests that all AGN in our classification have similar levels of extinction, with column densities on the order of $N_H\sim10^{22}-10^{23}$ \cite[see][]{2015ApJ...800L..10R}. This is inconsistent with a dominant budget of unobscured AGN light arising at UV wavelengths among the compact sources \cite[that could potentially influence rapid morphological transformations in the hosts, e.g.][]{2010MNRAS.405..718P}. Indeed, these $N_H$ values are in the typical range of obscuration observed for Type 2 AGN at $z=2$ \citep{2016ApJ...830..100M}.

\begin{figure*}
\centering
\includegraphics[width=0.40\textwidth]{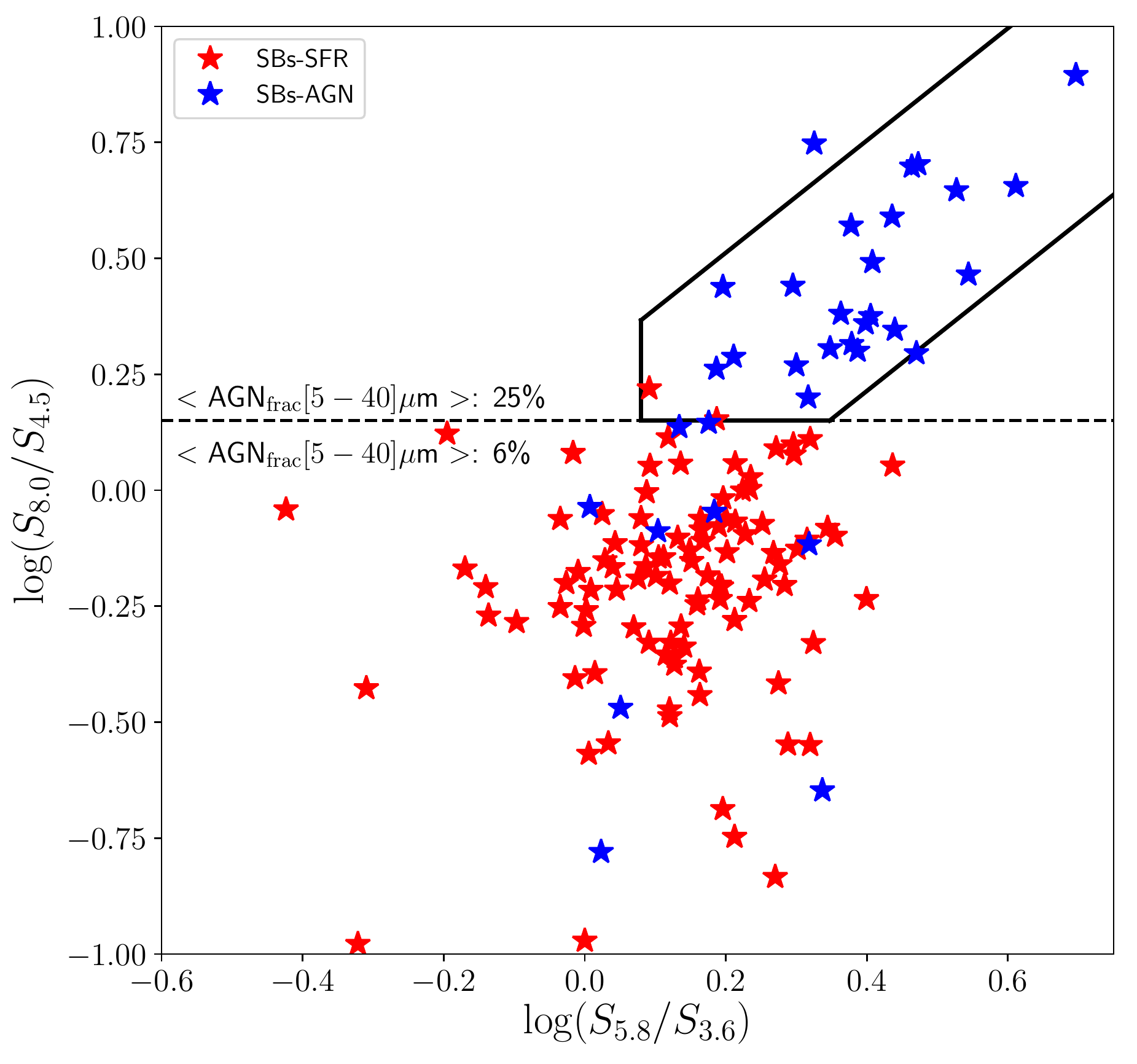}
\includegraphics[width=0.59\textwidth]{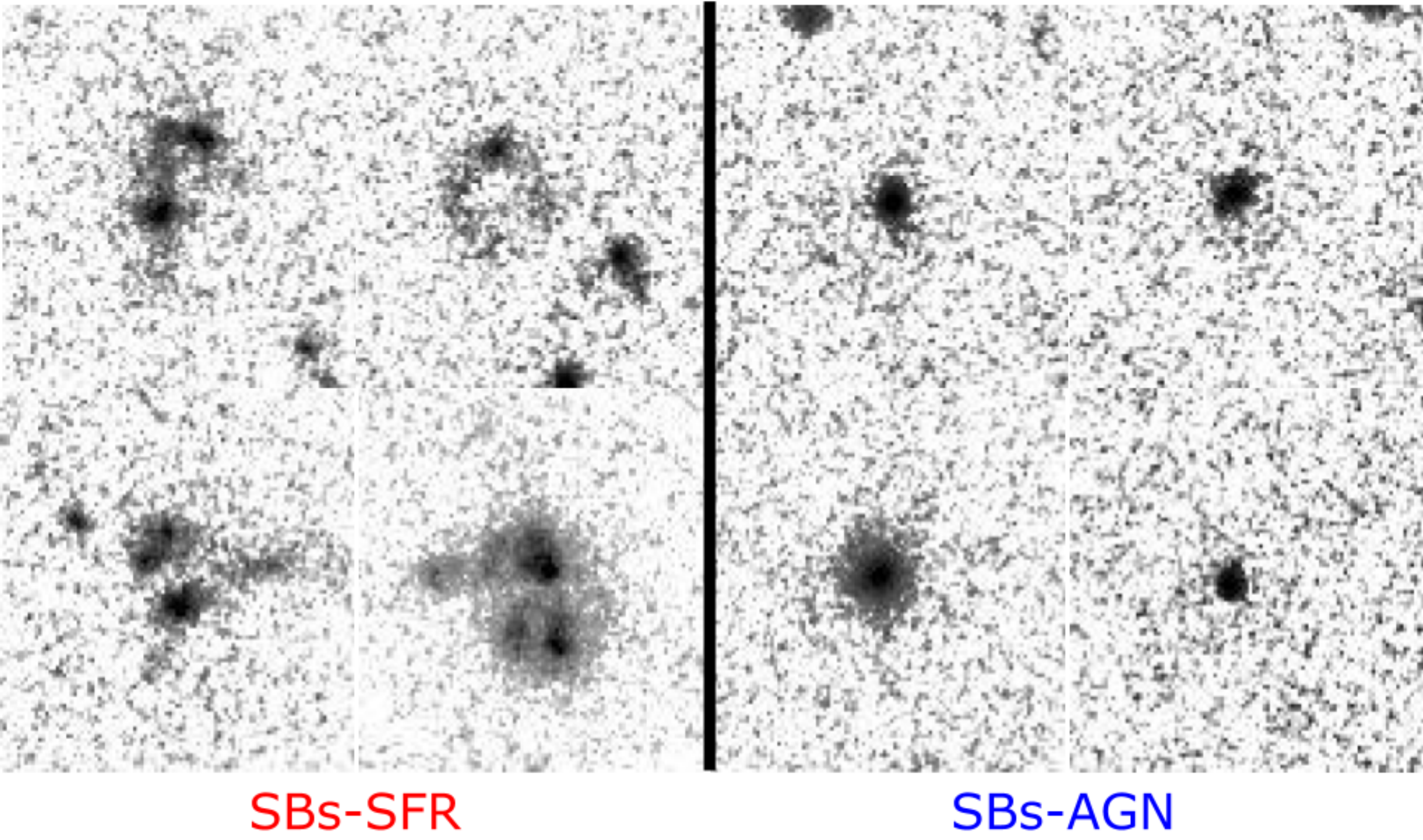}
\caption{
    {\it Left panel:} Colour-colour IRAC 
    plot of the 152 SBs selected in COSMOS. Blue stars are classified as SBs-AGN (35 in total), red stars as SBs-SFR (117). Black lines mark the position of the AGN selection criteria from \cite{2012ApJ...748..142D}. 
    The mean value of the sample AGN  contribution (in percentage units) between [5-40]$\mu$m over and below this value is reported.
    {\it Right panel:} Morphological examples of SFR dominated starbursts (left) and AGN dominated starbursts (right). HST-ACS cutouts have a size of 5" x 5".
    }
\label{fig:agn_cutout}
\end{figure*}

By including also the $i$-band undetected sources, the fraction of visual compact sources would be $\sim$11$\pm3$\% and $\sim$39$\pm7$\% for SBs-SFR and SBs-AGN, respectively.
Typical examples of the different classes are shown in Figure \ref{fig:agn_cutout}.
We caution that this morphological analysis is limited by the shallowness of the COSMOS/HST imaging, but also by the huge dust extinction associated to the SBs, that allows only a minor contribution of un-absorbed UV light to escape the obscured star-forming regions \citep{2017ApJ...838L..18P}. Despite these important limitations, we interpret our results as a supporting evidence that SBs-AGN are probably more compact and dense sources, likely corresponding to a final merger stage \cite[as many local ULIRGs,][]{1988ApJ...325...74S}.

\subsection{ALMA continuum counterparts}
We searched for possible ALMA millimetric counterparts of the SBs in the public archive, adopting a search radius of 2.5$"$.
The final sample includes 33 sources with a clear detection over $3\sigma$, mostly in band 7 at 870$\mu$m (with 4 un-detected sources that we do not include in this work), with few of them detected also in band 3 or 2.  The measurement sets were calibrated running the data reduction scripts with the Common Astronomy Software Applications (CASA). Fluxes and associated errors were evaluated with CASA, by fitting the emissions with a bi-dimensional Gaussian, and are reported in Table \ref{tab:my_label}. 
For the 33 ALMA detected sources, we included the sub-mm fluxes in the observed SED and we performed a second fitting run, as described in Section\,\ref{sec:sed}. The updated physical parameters are reported in Table \ref{tab:my_label}.

\begin{table*}
    \centering
    \scriptsize
    \label{tab:my_label}
    \begin{tabular}{cccccccccccccc}
    \hline
    ID & RA    & DEC       & $z$   & $\log L_{IR}$ & $\log L_{\rm AGN}$ & $\log M_*$   & $\log M_{gas}$	& AGN &	${\rm F}_{\rm 870\,\mu m}$ & ${\rm F}_{\rm 1.3mm}$ & ${\rm F}_{\rm 3.0mm}$ \\
    COSMOS15 & [deg] & [deg]     &       & [L$_\odot$] & [erg/s] & [M$_\odot$]&[M$_\odot$]&     &[mJy] &[mJy] & [mJy] \\  
    \hline
    182648  & 150.64316 & 1.558194  & 1.6887  & 11.115 &  45.595 & 10.4616   &  9.840    & yes   &  0.49$\pm$0.10 & -             & - \\
    221280  & 149.76807 & 1.617000  & 2.3220  & 11.185 &  43.945 & 10.4703   & 10.282    & yes   &  1.11$\pm$0.22 & -             & - \\
    244448  & 150.01202 & 1.652130  & (1.5180)& 10.645 &       - & 10.7505   & 10.472    &  no   &  2.01$\pm$0.25 & -             & - \\
    280968  & 149.79010 & 1.711870  & 1.7844  & 11.343 &       - & 10.4035   & 10.384    &  no   &  1.33$\pm$0.25 & -             & - \\
    323041  & 149.81653 & 1.779770  & 2.0933  & 11.244 &  44.135 & 10.5122   & 10.018    & yes   &  1.38$\pm$0.46 & -             & - \\
    349784  & 150.48938 & 1.821710  & 1.9693  & 11.100 &       - & 10.8425   & 10.628    &  no   &  1.59$\pm$0.34 & -             & - \\
    386956  & 150.34194 & 1.880208  & 2.2493  & 11.346 &       - & 10.6594   & 10.201    &  no   &  1.88$\pm$0.30 & -             & - \\
    505526  & 150.42101 & 2.068100  & 2.2684  & 10.666 &  45.245 & 11.0934   & 11.313    & yes   & 11.93$\pm$0.71 & -             & - \\
    506667  & 150.72984 & 2.071170  & 2.4433  & 11.200 &       - & 10.5955   & 10.590    &  no   &  2.23$\pm$0.77 & -             & - \\
    524710  & 149.76853 & 2.099614  & 2.1059  & 11.321 &       - & 10.4187   & 10.214    &  no   &  1.72$\pm$0.23 & -             & - \\
    571598  & 150.61642 & 2.167971  & 1.5052  & 11.167 &       - & 10.8600   & 10.792    &  no   &  5.39$\pm$0.23 & -             & - \\
    578926  & 150.40103 & 2.180390  & (2.3341)& 11.567 &       - & 10.9570   & 10.617    &  no   &  2.05$\pm$0.55 & -             & - \\
    600601  & 150.13265 & 2.211946  & 1.9875  & 10.752 &  45.605 & 11.1349   & 11.081    & yes   &  8.27$\pm$0.44 & 2.46$\pm$0.10 & - \\
    605409  & 149.76813 & 2.219876  & (1.7766)& 11.112 &       - & 10.8708   & 10.585    &  no   &  3.78$\pm$1.03 & -             & - \\
    640026  & 150.03663 & 2.270976  & 1.7977  & 11.313 &  44.185 & 10.2632   & 10.349    & yes   &  1.09$\pm$0.27 & -             & - \\
    642313  & 149.60419 & 2.275064  & 2.0069  & 11.254 &       - & 10.6391   & 10.677    &  no   &  1.77$\pm$0.31 & -             & - \\
    651584  & 149.92196 & 2.289929  & 2.3341  & 11.600 &       - & 10.8786   & 10.626    &  no   &  4.93$\pm$0.34 & -             & - \\
    734578  & 149.52823 & 2.413200  & 1.9641  & 11.765 &       - & 10.5523   & 10.893    &  no   &  -             & 1.88$\pm$0.23 & - \\
    745498  & 150.46551 & 2.429549  & 1.6332  & 10.731 &       - & 10.7144   & 10.355    &  no   &  3.47$\pm$0.34 & -             & - \\
    747590  & 150.22447 & 2.433010  & 1.6351  & 10.903 &       - & 10.6513   & 10.285    &  no   &  1.60$\pm$0.24 & -             & - \\
    752016  & 150.33683 & 2.439920  & 2.2682  & 11.585 &       - & 10.7232   & 10.665    &  no   &  4.50$\pm$0.31 & -             & - \\
    754372  & 150.06907 & 2.444010  & 2.4355  & 11.439 &  45.595 & 11.0108   & 10.468    & yes   &  4.86$\pm$0.65 & -   & 0.275$\pm$0.065 \\
    769248  & 150.25528 & 2.466839  & (2.2640)& 10.903 &       - & 10.5134   & 10.629    &  no   &  3.80$\pm$0.42 & -             & - \\
    794848  & 150.09341 & 2.507339  & 2.1990  & 11.365 &  45.975 & 10.7362   & 10.648    & yes   &  3.09$\pm$0.25 & -             & - \\
    810228  & 150.11307 & 2.528020  & 2.0167  & 11.264 &       - & 10.6265   & 10.687    &  no   &  -             & -   & 0.137$\pm$0.060 \\ 
    815012  & 150.60329 & 2.536536  & (2.2872)& 11.119 &       - & 11.1034   & 10.692    &  no   &  6.70$\pm$0.35 & -             & - \\ 
    818426  & 150.72202 & 2.541904  & 2.2664  & 11.607 &  45.935 & 10.6405   & 10.807    & yes   &  1.01$\pm$0.30 & -             & - \\ 
    842595  & 149.99796 & 2.578227  & 2.4200  & 11.693 &  44.205 & 10.7809   & 10.732    & yes   &  1.97$\pm$0.32 & -             & - \\ 
    902320  & 150.03726 & 2.669600  & (1.5990)& 10.274 &       - & 10.9826   & 11.162    &  no   &  6.41$\pm$0.39 & -   & 0.147$\pm$0.057 \\ 
    917423  & 149.99218 & 2.693436  & 2.1284  & 11.330 &       - & 10.7933   & 10.581    &  no   &  1.77$\pm$0.34 & -             & - \\ 
    917546  & 150.16165 & 2.691588  & (1.9745)& 10.937 &  44.785 & 10.4437   & 10.156    & yes   &  1.39$\pm$0.25 & -             & - \\ 
    951838  & 150.26832 & 2.749270  & 2.0186  & 11.128 &  45.535 &  9.9814   & 10.076    & yes   &  1.60$\pm$0.24 & -             & - \\       
    980250  & 150.01611 & 2.792355  & 1.7598  & 10.533 &       - & 11.1303   & 10.875    &  no   &  5.03$\pm$0.25 & -             & - \\

    \hline
    \end{tabular}
\end{table*}

\begin{figure*}
\centering
\includegraphics[width=1.\textwidth]{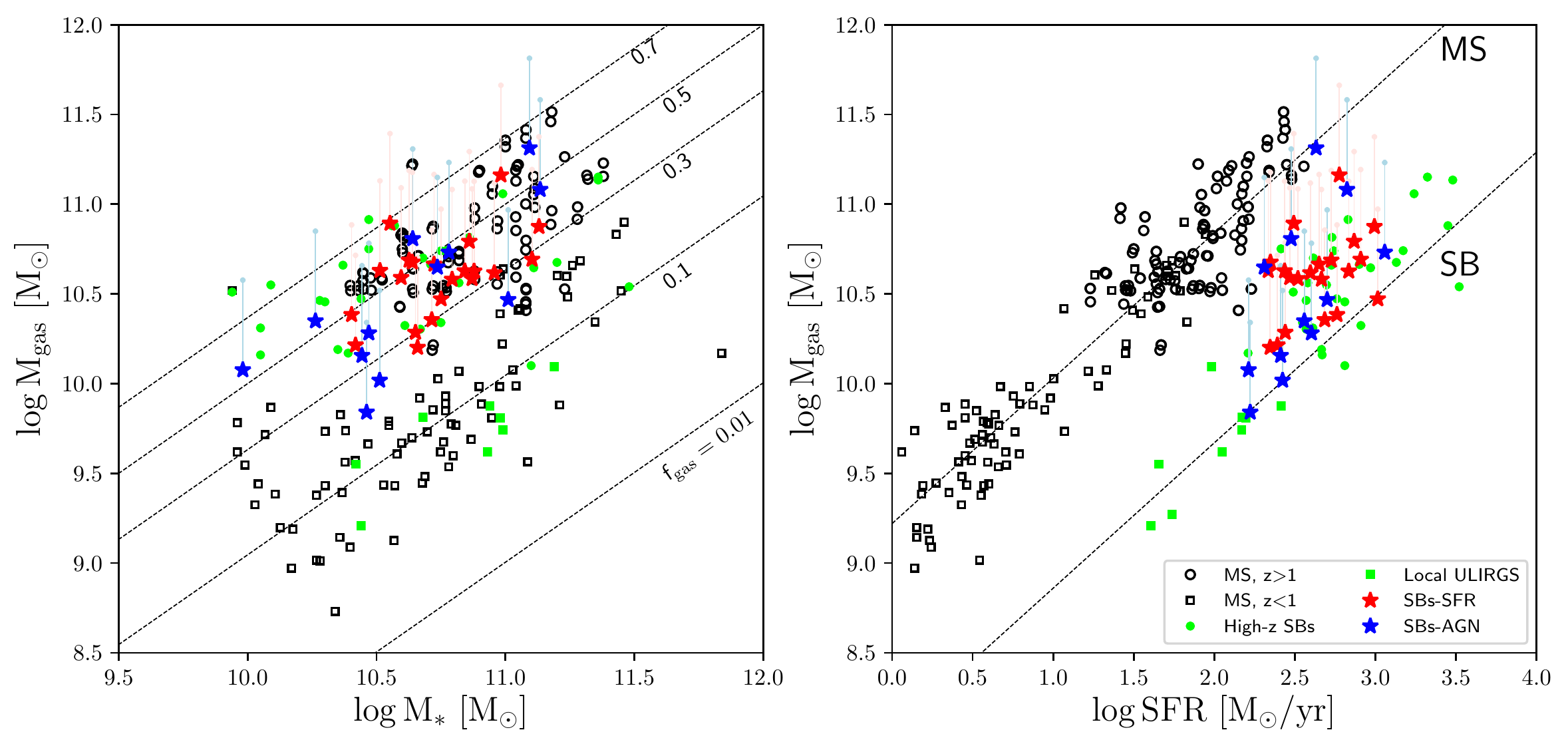}
\caption{
    {\it Left panel:} $M_{gas}$ versus $M_*$ for the SBs with an ALMA detection (blue stars: SBs-AGN; red stars: SBs-SFR). $M_{gas}$ are evaluated adopting a solar metallicity and an $\alpha_{CO}$ of 0.8, corresponding to a DGR of 30, 
    and represent our reference values; light points and relative connectors still refer to solar metallicity, but with $\alpha_{CO} = 1.0$. Formal errors are smaller then the difference computed with the two methods. Loci of constant gas fraction (i.e. $M_{gas}/(M_{gas}+M_*)$), are shown with dotted lines. Open black squares (circles) represent MS galaxies at $z < 1 (> 1)$ as compiled by \cite{2014ApJ...793...19S} and \cite{2018arXiv180703378P}. Green symbols show local ULIRGs (filled squares) and high-$z$ starbursts (filled circles). {\it Right panel:} $M_{gas}$ versus $SFR$. 
    Empirical model curves (dashed lines) for MS and ULIRGs/SB are described in \cite{2014ApJ...793...19S}. 
}
\label{fig:mgas}
\end{figure*}


\section{Results}
In this Section, we report the main results of our analysis of the gas masses computed for the SBs with an ALMA continuum detection. Out of 33 sources, 12 objects turn to be SBs-AGN, thus representing 36$\%$ of our sample. We note that this limited ALMA detected data-set is quite representative of the whole SB population in this redshift range (in terms of $M_*$, elevation above the MS and AGN content, see Figure \ref{fig:selection}). 

\subsection{Comparison sample}
For comparative analyses, we use the reference sample  compiled by \cite{2014ApJ...793...19S} and \cite{2018arXiv180703378P},  including 'typical' star-forming galaxies at $z\leq3$ with measurements of their CO luminosity. Local ULIRGs and high-$z$ starbursts with a CO detection are added. 
Detailed references for the various samples included are reported in \cite{2014ApJ...793...19S} and \cite{2018arXiv180703378P}. 
We added the recent compilation of SBs by \cite{2015ApJ...812L..23S} and Silverman et al. (2018), at $z\sim1.6$.
\begin{figure*}
\centering
\includegraphics[width=.8\textwidth]{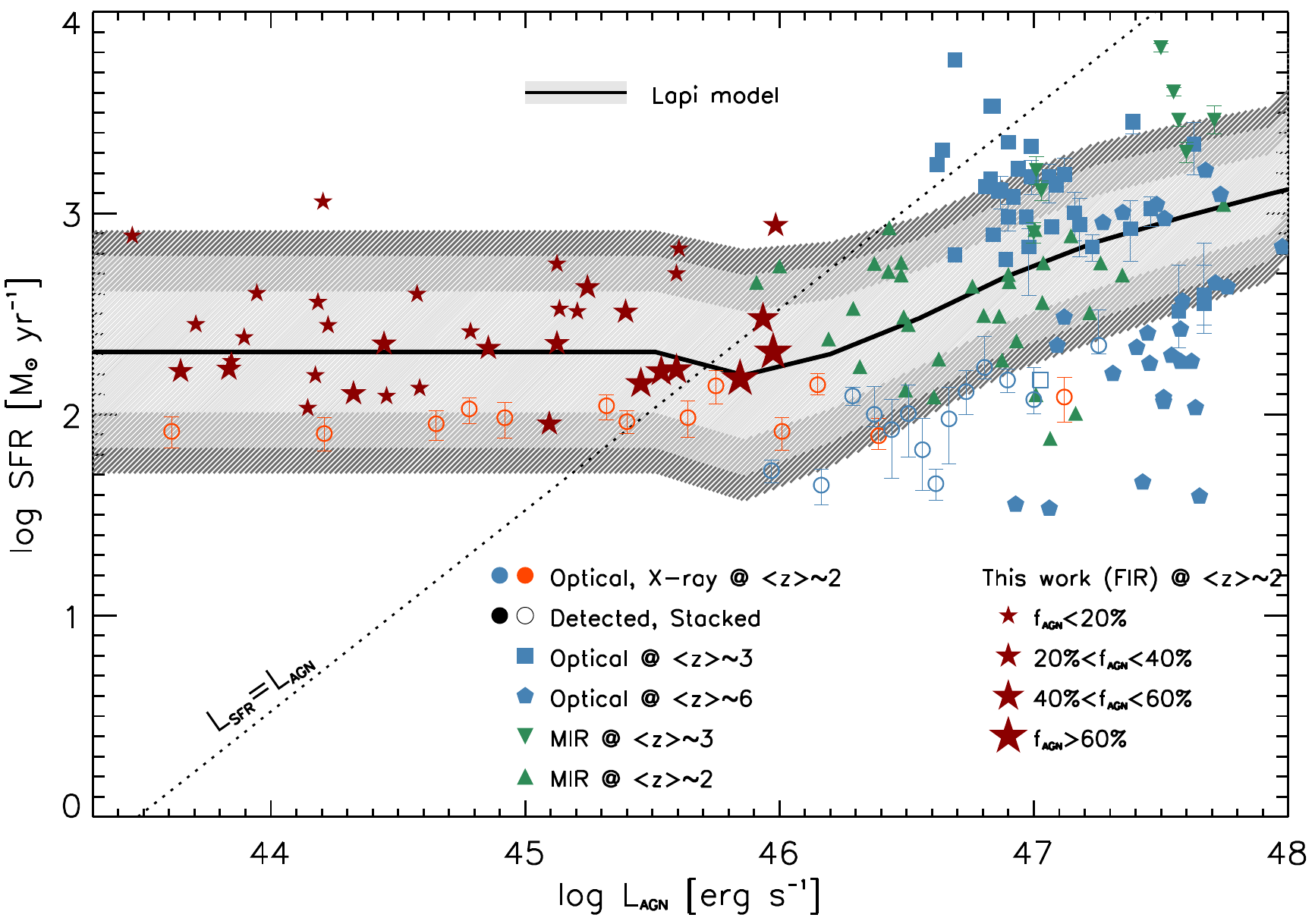}
\caption{
    Co-evolution plane between bolometric AGN luminosity and SFR in the hosts $z>1.5$. 
    The gray shaded areas show the average locus of the evolutionary tracks from the model of \cite{2018ApJ...857...22L}, adopting a SFR detection threshold around $10^2\, M_\odot$ yr$^{-1}$. Orange symbols refer to X-ray AGNs, green to mid-IR selected AGNs, blue to optically selected QSO. Complete references are reported in \cite{2018arXiv180606516B}. We include all SBs-AGN from our original sample (red filled stars):  size of the symbol is proportional to the AGN luminosity contribution in the 5-40 $\mu$m range, derived from $SED3FIT$. SFR for the SBs in our sample is derived from $L_{IR}$ in Table 1.
}
\label{fig:lapimodel}
\end{figure*}

\subsection{Dust and molecular gas masses}
\label{mgas}
Dust masses ($M_{dust}$) have been derived following the procedure described in \cite{2012ApJ...760....6M, 2017A&A...603A..93M} by fitting the SED in the IRAC-ALMA observed frame with the \cite{2007ApJ...657..810D} models. The total gas mass ($M_{gas}$, which incorporates both the molecular and atomic phases) can be inferred from the dust mass through the metallicity dependent gas-to-dust ratio \citep[{\it GDR}, e.g.][]{2010A&A...518L..62E, 2012ApJ...760....6M}: $M_{gas} = M_{dust}/GDR(Z)$.
For consistency with $M_{gas}$ estimates of SBs in the literature, we adopt  $GDR \approx 30$, that based on \cite{2012ApJ...760....6M} corresponds to a CO-to-$M_{gas}$ $\alpha_{co}$ conversion factor of 0.8 M$_{\odot}$ / (K km s$^{-1}$ pc$^{2}$), typically used for strong starbursts. For completeness we also infer $M_{gas}$ estimates with $GDR \approx 95$ that corresponds to solar gas-phase metallicity that could be considered as a lower limit {\bf (on metallicity, hence  upper limits on $GDR$ and thus on $M_{gas}$)} for dust obscured SBs  \cite[e.g.][]{2017ApJ...838L..18P}. These $M_{gas}$ estimates are consistent (within 0.15dex) with the average $M_{gas}$ estimates derived based on the monochromatic flux densities in the R-J tail of the SED (one or more of 870, 1300, and 3000\,$\mu$m in our case) and the recipe presented in  \cite{2017ApJ...837..150S}.
 
\subsection{Gas masses and gas fractions in SBs-AGN and SBs-SFR}
We report in Figure \ref{fig:mgas} (left panel) the gas masses of the SBs as a function of their $M_*$, divided into  SBs-AGN and SBs-SFR (blue and red filled stars, respectively). The SB population is dominated by gas rich galaxies, with gas fractions (defined as $f_{gas}=M_{gas}/(M_{gas}+M_*)$) spanning the range 20\%-70\%. This is similar to the typical $f_{gas}$ ($\sim50\%$) of the normal star-forming sources at similar $z$ (open circles).
Local star-forming galaxies (both MS and ULIRGs/SBs) are instead much gas poorer, with $f_{gas}\sim10\%$, as expected on the basis of the observed gas fraction evolution with cosmic time \cite[e.g.][and references therein]{2012ApJ...758L...9M, 2018ApJ...853..179T}.\\
When looking at the separate gas fractions of SBs-AGN and SBs-SFR, we do not observe a significant difference, providing average values of $f_{gas}=43\,\pm\,4\%$ and $f_{gas}=42\,\pm\,2\%$ for the two classes, respectively. 

To shed more light on our analysis in the most efficient star-forming sources, we study (right panel of Figure\,\ref{fig:mgas}) the distribution of $M_{gas}$ as a function of SFR, in order to compare the star formation efficiency (i.e. SFE = SFR/$M_{gas}$) of SBs-AGN and SBs-SFR. Our sample turns to be "SFR-selected" by construction, with SFR $\ge150M_{\odot}/yr$ due to the requirement of being $Herschel$ selected (see Figure\,\ref{fig:selection}). We then observe that SBs lie on a contiguous sequence of increasing SFE, that fills the gap between the two paradigmatic sequences of normal galaxies and ULIRG/SBs \cite[dotted lines, as from][]{2014ApJ...793...19S}, usually interpreted as the main loci of two extremely different SF modes. Our results support recent works suggesting  the existence of a continuous increase in SFE with elevation from the main sequence, as opposed to a bimodal distribution \citep{2018ApJ...867...92S}.

\section{Discussion}

\subsection{Comparison with the merger triggered SB-QSO evolutionary sequence}
As mentioned in Section \ref{sec:intro}, we can compare our results with the expectations of the AGN-galaxy co-evolutionary scenario, that predicts a luminous IR phase of buried SMBH growth, co-existing with a starburst \cite[likely arising from a merger,][]{2018Natur.563..214K, 2018ApJ...853...63D} before feedback phenomena deplete the cold molecular gas reservoir of the galaxy and an optically luminous quasar (QSO) shines out \citep{2008ApJS..175..356H}.
On one side, we have qualitatively observed that starbursts including an AGN have on average more compact and nucleated UV-restframe morphologies with respect to "inactive" SBs, suggesting that they are kept in a different dynamical evolutionary phase. This could correspond to the key transition when the late mergers trigger a high SFR, before the fully developed AGN phase. Simulations and observations, indeed, suggests a temporal delay between the peak of the SFR and the peak of the BHAR \citep{2015ApJ...800L..10R, 2016A&A...587A..72B, 2016ilgp.confE...7L}.
On the other side, we did not find a significant reduction of gas fractions in the SBs-AGN hosts compared to "inactive" SBs-SFR. We argue that, if major mergers are the main triggering mechanism of obscured BHAR in SBs, feedback phenomena (producing large outflows from the central BH) are not efficient in removing significant amount of molecular gas in the host galaxies.

\subsection{Starbursts as primordial galaxies}
An alternative interpretation for the properties of the off-MS galaxies is provided by an in-situ co-evolution scenario \citep{2018ApJ...857...22L}, 
envisaging high-$z$ galaxy evolution to be mainly ruled by the interplay between in-situ processes like gas infall, compaction, star formation, accretion onto the hosted central SMBHs, and related feedback processes (with wet galaxy mergers having a minor role at $z\geqslant 1$). In this framework $z\sim 2$ sources with SFR $\geqslant$ a few $10^2\, M_{\odot}$ constitute the progenitor of local massive spheroids with stellar mass $M_* \geqslant 10^{11}\, M_{\odot}$. During their star-forming phase, lasting some $10^8$ yr, these galaxies feature a nearly constant SFR and a linearly increasing stellar mass. In the SFR-$M_*$ diagram they follow an almost horizontal track (see cyan tracks in Fig.\,\ref{fig:selection}) while moving toward the galaxy MS locus;  there they will have acquired most of their mass before being quenched by energy/momentum feedback from the SMBH. 

Being in the early stages of their evolution, the SBs can host only a rather small SMBH, originating a bolometric AGN luminosity $L_{AGN}$ weaker than that $L_{SFR}$ associated to the SFR in the host. However, the BH mass is expected to grow exponentially 
generating a noticeable statistical variance in $L_{AGN}$ at given SFR. As a consequence, in the SFR vs. $L_{AGN}$ plane, SBs are expected to populate a strip parallel to the $L_{AGN}$ axis, and located to the left of the locus where $L_{SFR} = L_{AGN}$ \cite[see Figure 4 and ][]{2018arXiv180606516B}.
The hosted AGNs are expected to be obscured and
with luminosity still not powerful enough to originate substantial feedback effects on the ISM of the host galaxy; thus, the SFR and the gas mass of the host are still not much affected. 
The compact morphologies of SBs-AGN (possibly linked to a forming bulge) could indicate that these sources are observed when the host stellar mass and the BH mass are reaching their maximum, just before  the feedback gets into action and  the BHAR and SFR gets quenched \citep{2014ApJ...782...69L}; further size evolution of the stellar component may be induced by the feedback itself and by late-time mass addition from dry mergers \cite[see][]{2018ApJ...857...22L}.


\section{Summary}
In conclusion, the results presented in this work are consistent with the idea that the SB population could be filled by a mixture of: 1) a class of highly star-forming merging sources (preferentially among the SBs-SFR), and 2) primordial galaxies, quickly accreting their $M_*$ together with their BH (mainly the SBs-AGN). 
If the level of AGN luminosity (proportional to BHAR) is also  correlated with the power of feedback, then IR-selected SBs are mainly low-luminosity AGN, and feedback effects have not yet started to eventually reduce the gas fraction of the objects \cite[as observed for local Seyfert in a similar AGN luminosity range,][]{2018MNRAS.473.5658R}.
As time passes, the galaxies will get older, the luminosity of the AGN will overwhelm that of the SFR in the host (moving toward the bottom right region of Fig.\,\ref{fig:lapimodel}), removing gas and dust from the ISM and quenching the SFR and the accretion itself. The systems will then shine as a bright optical quasar (blue symbols in Fig.\,\ref{fig:lapimodel}), before turning into passive sources.
To better constrain the dominant nature of starbursts (mergers versus in-situ formation and evolution), much larger samples are required to provide a statistical description of the gas content in this population and to understand the impact of the co-evolving obscured central activity. An important discriminating factor will be provided by future spatially resolved  kinematic studies of the stellar or gas components ($JWST$, $ALMA$), currently available for just few sources \cite[e.g.][]{2018ApJ...867...92S}.

\acknowledgments
GR, CM, AR and AP acknowledge support from PRIN-SKA ESKAPE-HI (PI L.Hunt) and from the 
STARS@UniPD grant.
GR and AL are supported by PRIN MIUR 2017 prot. 20173ML3WW\_002
“Opening the ALMA window on the cosmic evolution of gas, stars and supermassive black holes“.
MP acknowledges support from the ESO Scientific Visitor Programme.
  GEM acknowledges support from the Villum Fonden research grant 13160, 
the Cosmic Dawn Center of Excellence,
the ERC
Consolidator Grant funding scheme (project ConTExt, grant number 648179).
We thank Marcella Brusa for fruitful discussions.

\end{document}